\documentclass [twocolumn,final]{svjour3}

\usepackage{amsmath}
\usepackage{amssymb}
\usepackage{graphicx}

\usepackage[numbers]{natbib}   
\newcommand{\vect}[1]{\vec{#1}}    
\newcommand{\ELL}{\mathfrak{L}}
\newcommand{\bb}[1]{\boldsymbol {#1}}
\newcommand{\II}[1]{\mathcal{#1}}

\newcommand{\ti}[1]{#1}  
\newcommand{\vol}[1]{\bf #1}  

\journalname{Climate Dynamics}

\allowdisplaybreaks[4]

\begin{document}


\title{Maximum-Entropy Weighting of Multiple Earth Climate Models}

\author{Robert K. Niven}

\institute{Robert K. Niven \at
              School of Engineering and Information Technology, The University of New South Wales at ADFA, Northcott Drive, Canberra, ACT, 2600, Australia. 
              \and Currently at Institut Pprime, CNRS / UniversitŽ de Poitiers / ENSMA, CEAT, 43 rue de l'AŽrodrome, F-86036 Poitiers Cedex, France.
              \\
              \email{r.niven@adfa.edu.au}      
}


\date{8 April 2011; revised 30 June 2011; accepted 8 August 2011}




\maketitle

\begin{abstract} 
A maximum entropy-based framework is presented for the synthesis of projections from multiple Earth climate models. {This identifies the most representative (most probable) model from a set of climate models -- as defined by specified constraints -- eliminating the need to calculate the entire set. Two approaches are developed, based on individual climate models or ensembles of models, subject to a single cost (energy) constraint or competing cost-benefit constraints. A finite-time limit on the minimum cost of modifying a model synthesis framework, at finite rates of change, is also reported. }


\end{abstract}

\section{\label{intro}Introduction}
A major challenge facing humanity is the possibility of climate change due to human and/or natural forcings, and how best to respond in a rational and informed manner. To this end, detailed global circulation models (GCMs) have been developed to predict the behaviour of the Earth climate system (atmosphere and oceans), involving solution of the continuity, Navier-Stokes, angular momentum and energy equations and constitutive relations over two- or three-dimensional domains, subject to various initial and boundary conditions \cite{Peixoto_O_1992, McGuffie_HS_2005}. These are run interrogatively to {yield {\it climate projections} -- predictions as a function of future time -- to examine various forcing and response scenarios.}  However, a serious difficulty for policy-makers is the promulgation of multiple models by different research groups, due to different modelling priorities, assumptions and input parameters, and inherent difficulties in the construction of climate models, especially in the handling of {coupled phenomena (e.g. humidity \cite{Paltridge_etal_2009}) and the need to dramatically reduce their computational complexity}, 
necessitating a turbulence closure scheme.  Even with the same (or similar) inputs, different models can provide significantly different climate projections \cite{IPCC_2010}.  A rational framework for the synthesis of such projections -- which operates in a transparent and fully defensible manner -- is urgently required, {to avoid the lack of objectivity of seemingly {\it ad hoc} amalgamations of projections from different groups.} 

Over the past century, {\it maximum entropy} (MaxEnt) methods have been developed for the construction of probabilistic models, initially in thermodynamics \cite{Boltzmann_1877, Planck_1901} and subsequently for all probabilistic systems \cite{Jaynes_1957, Jaynes_1963, Jaynes_2003}. Although imbued with several information-theoretic interpretations \cite{Jaynes_1957, Jaynes_1963, Jaynes_2003, Shannon_1948}, the success of such models rests ultimately on the {\it maximum probability} (MaxProb) principle of Boltzmann \cite{Boltzmann_1877, Planck_1901, Vincze_1972, Grendar_G_2001, Niven_2005, Niven_2006, Niven_2009a, Niven_2009b, Niven_G_2009, Grendar_N_2010}: 
``{\it a system can be represented by its most probable state.}''  This provides a probabilistic definition of the (relative) entropy function:
\begin{gather}
\mathfrak{H}_{rel} = K \ln \mathbb{P}
\label{eq:Boltz}
\end{gather}
where $\mathbb{P}$ is the governing probability of an observable realization (macrostate) of the system and $K$ is a constant. The maximum of $\mathfrak{H}_{rel}$ thus coincides with that of $\mathbb{P}$.  If the system can be represented by the allocation of distinguishable balls (objects) to distinguishable boxes (categories), then $\mathbb{P}$ will satisfy the multinomial distribution \cite{Brillouin_1930, Fortet_1977, Read_1983}:
\begin{gather}
\mathbb{P}=\text{Prob}(n_1,...,n_s|q_1,...,q_s,N) =N! \prod\limits_{i = 1}^s \frac{q_{i}^{n_{i}} }{n_{i}!}
\label{eq:multinomial}
\end{gather}
where $n_i$ is the observed occupancy (number of balls) and $q_i$ is the prior probability of the $i$th category, $N$ is the total number of balls and $s$ the number of categories.  Insertion of \eqref{eq:multinomial} into \eqref{eq:Boltz} with $K=N^{-1}$, taking the asymptotic limit $N \to \infty$ and $n_i/N \to p_i$, gives the (negative) Kullback-Leibler entropy function:
\begin{gather}
\mathfrak{H}_{KL} = -\sum\limits_{i=1}^s p_i \ln \frac{p_i}{q_i}
\label{eq:KL}
\end{gather}
Maximisation of \eqref{eq:KL} for a system {which satisfies \eqref{eq:multinomial},} subject to its constraints, is therefore equivalent to seeking its most probable realization in the asymptotic limit, subject to the same constraints.  


{
We therefore adopt a broader concept of `entropy' than that normally used in the physical sciences.  Climatologists will be familiar with the {\it thermodynamic entropy} $S$, which has a clearly defined meaning as the state function $S = \int \delta Q/T$ (Clausius) or $S = k \ln \mathbb{W}$ (Boltzmann), where $\delta Q$ is the increment of heat entering a system, $T$ is temperature, $k$ is the Boltzmann constant and $\mathbb{W}$ is the number of microstates within a given realization (macrostate) of a system. Its rate of change is $dS/dt$, of which the excess (exported) component is commonly termed the {\it thermodynamic entropy production} $\dot{\sigma}$ \cite{deGroot_M_1962, Niven_2009c, Niven_2010}.  However, under the MaxProb or MaxEnt approach adopted here, entropy acquires a more fundamental meaning in terms of the probabilistic state space of a system, however defined. To emphasise their generic character, such entropies are here denoted $\mathfrak{H}$.  The thermodynamic entropy is in fact a special case of the generic, being derivable by the application of MaxEnt to an energetic system \cite{Boltzmann_1877, Planck_1901, Jaynes_1957, Jaynes_1963, Jaynes_2003}. The ensuing analyses are based entirely on generic entropy functions, not necessarily related to $S$; that said, much of the underlying mathematical structure is identical.}

The aim of this study is to construct a framework for the synthesis of climate projections from multiple climate models, based on the MaxProb (hence MaxEnt) principle. By analogy with thermodynamics, two approaches are presented, involving constraints on the properties of individual climate models or of ensembles of climate models.  {In each case, the analysis identifies the most representative (most probable) model from a set of climate models, circumventing the need to calculate the entire set.  Other} implications of these frameworks, which arise from the mathematical structure given by Jaynes \cite{Jaynes_1957, Jaynes_1963, Jaynes_2003}, are examined. In addition, we report a curious finite-time limit on the minimum cost of varying the overall framework at specified rates of change, using a theorem from finite-time thermodynamics \cite{Weinhold_1975c, Ruppeiner_1979, Salamon_A_G_B_1980, Salamon_B_1983, 
Nulton_etal_1985, 
Niven_Andresen_2009}.

\section{\label{deriv}Derivations}
\begin{figure}[t] 
\setlength{\unitlength}{0.6pt}
  \begin{picture}(410,230)
   \put(10,10){\includegraphics[width=80mm]{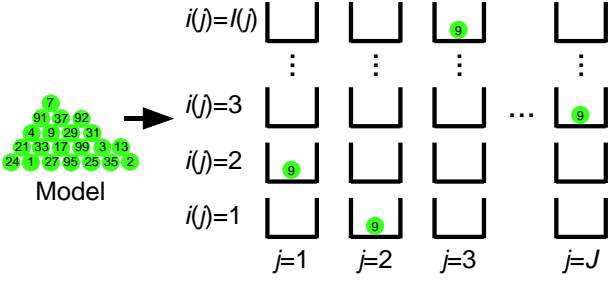} }
  \end{picture}
\caption{Generic combinatorial representation of the climate model weighting framework, showing a single model composed of individual discrete choices of the $i(j)$th algorithm or methodology for each model component $j=1,...,J$.} 
\label{fig:comb1}
\end{figure}

%
Consider an individual Earth general climate model (GCM), composed of $J$ separable computational components. Each component $j = 1,..., J$ is executed by a single choice $i(j)$ of algorithm, methodology or paradigm, from a total of $I(j)$ possible choices. As shown in Figure \ref{fig:comb1}, this gives a combinatorial scheme in which an individual model is constructed from a set of unique choices $i(j) \in \{1,...,I(j)\}, \forall j$.  We assume that all models are calculated using the same set of input parameters and assumptions $\vect{\theta}$; moreover, to accommodate variability or errors in $\vect{\theta}$, each model will yield a set or domain of climate projections, which can be explored by Monte Carlo analysis or by some other means.  
If we move beyond the deterministic mindset that an individual climate model must be the ``correct'' one, how should we weight the projection sets from different climate models, to obtain a (statistical) picture of their merged sets of projections?  One could simply combine an available set of model outputs using equal or assigned weighting factors, as suggested in \cite{IPCC_2010}, but unless every possible combination has been computed, the resulting composite model will be rather arbitrary. In addition, if the model space 
is infinite (or merely very large), it will be impossible to compute the composite model in the lifetime of the universe (or in any reasonable time frame).  Moreover, the use of equal weights does not allow the incorporation of additional constraints on the model space.  We therefore propose a MaxProb-based (hence MaxEnt-based) framework for the weighting of multiple climate models, for which two distinct approaches are available.

\subsection{\label{modelA} ``Microcanonical'' Framework}

\begin{figure}[t] 
\setlength{\unitlength}{0.6pt}
  \begin{picture}(410,460)
   \put(0,0){\small (b)}
   \put(10,10){\includegraphics[width=80mm]{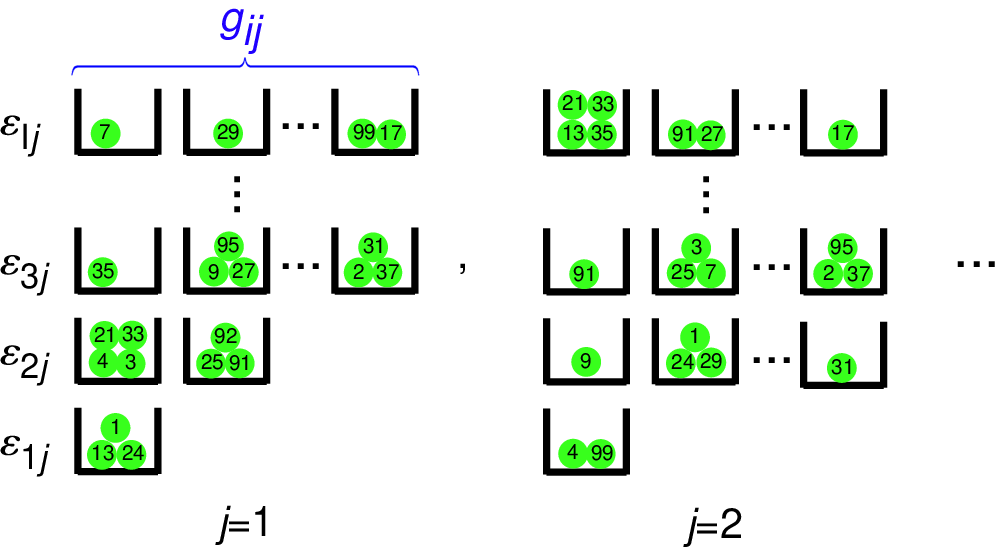} }
   \put(0,230){\small (a)}
   \put(10,240){\includegraphics[width=80mm]{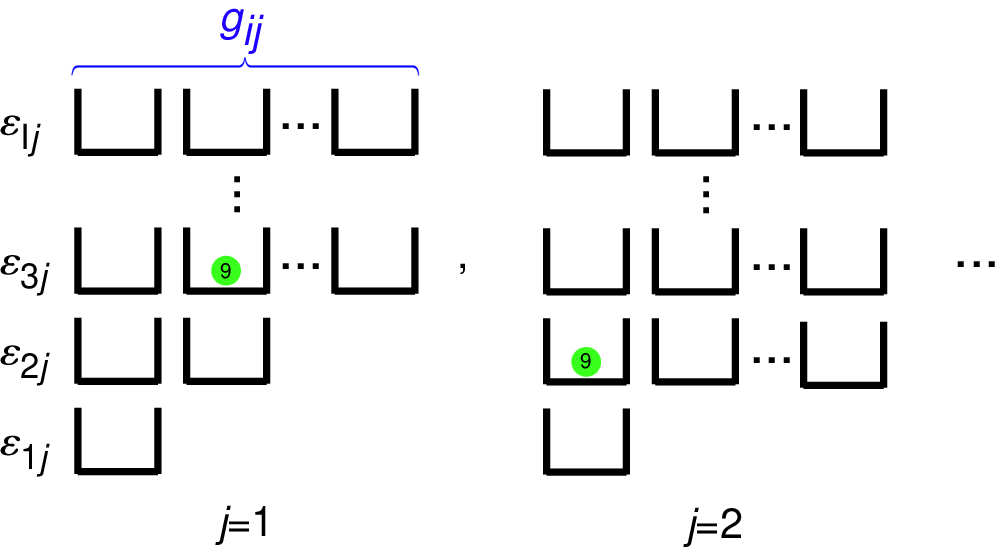} }
  \end{picture}
\caption{Combinatorial representations of (a) the microcanonical framework, showing a single model composed of $g_{ij}$ degenerate choices $i(j)$ for each model component $j=1,...,J$ (ranked by energy level $\epsilon_{ij}$); and (b) the canonical framework, composed of an ensemble of $N$ amalgamated microcanonical models. Ball numbers denote the model index. }
\label{fig:comb2}
\end{figure}
We first construct a ``microcanonical'' climate model weighting framework, based on the properties of individual climate models. Extending the representation in Figure \ref{fig:comb1}, consider a single climate model shown in Figure \ref{fig:comb2}(a), in which we choose to rank each choice of algorithm or method $i(j)$ by its cost or energy $\epsilon_{ij}$, indicating (for example) the relative programming and computational cost of execution of this particular choice. Each energy level $i(j)$ is considered to have the degeneracy $g_{ij} \ge 1$, equal to the number of choices which share the same cost $\epsilon_{ij}$. The ranking scheme $i(j)$ therefore accounts for, but does not distinguish between, choices of equal cost.  Each level $i(j)$ is taken to have the occupancy $m_{ij} \in \{0,1\}$ (the choices are unique).  From simple probabilistic considerations \citep{Brillouin_1930, Fortet_1977, Read_1983} for equiprobable degenerate choices, the probability of a given choice $i(j)$ is given by the reduced multinomial distribution:
\begin{equation}
\mathbb{P}_{i|j}^{(\mu)} = \text{Prob}(\vect{m}_j|\vect{g}_j,\vect{\theta}) = \frac{1}{G_j} \prod\limits_{i=1}^{I(j)} \frac{g_{ij}^{m_{ij}}}{m_{ij}!} 
\label{eq:Pj_micro}
\end{equation}
where $\vect{m}_j = \{m_{1j},...,m_{I(j)j}\}$, $\vect{g}_j= \{g_{1j},...,g_{I(j)j}\}$, $G_j = \sum\nolimits_{i=1}^{I(j)} g_{ij}$ and superscript $\mu$ denotes the microcanonical framework. Eq.\ \eqref{eq:Pj_micro} reduces to $\mathbb{P}_{\tau|j}^{(\mu)}  = {g_{\tau j}}/{G_j}$, where $\tau(j)$ is the selected choice, but it is preferable to keep the $m_{ij}$ explicit using \eqref{eq:Pj_micro}. The probability of selecting a single overall model, assuming that the $J$ components are independent, is therefore given by the ``multi-multinomial'':
\begin{equation}
\begin{split}
\mathbb{P}^{(\mu)} &= \text{Prob}(\vect{m}|\vect{g},\vect{\theta})  = \prod\limits_{j=1}^{J} \text{Prob}(\vect{m}_j|\vect{g}_j,\vect{\theta}) 
\\
&= \prod\limits_{j=1}^{J} \mathbb{P}_{i|j}^{(\mu)} = \prod\limits_{j=1}^{J} \frac{1}{G_j} \prod\limits_{i=1}^{I(j)} \frac{g_{ij}^{m_{ij}}}{m_{ij}!} 
\end{split}
\label{eq:P_micro}
\end{equation}
where $\vect{m}$ and $\vect{g}$ are the respective matrices of $m_{ij}$ and $g_{ij}$. Each model is subject to $J$ constraints on the total occupancy within each component $j$:
\begin{gather}
\sum\limits_{i=1}^{I(j)} m_{ij} = 1, \forall j
\label{eq:C0j_micro}
\end{gather}
Assuming that the costs are additive over the $J$ components, we can also include a constraint on the total cost $E$ of running the overall model:
\begin{gather}
\sum\limits_{j=1}^J \sum\limits_{i=1}^{I(j)} \epsilon_{ij} m_{ij} = E
\label{eq:CE_micro}
\end{gather}
To determine the most probable or {\it equilibrium} 
model, given the above occupancy and total energy constraints, we should maximise \eqref{eq:P_micro} with respect to the unknowns $\{m_{ij}\}$, subject to \eqref{eq:C0j_micro}-\eqref{eq:CE_micro}. From the Boltzmann definition \eqref{eq:Boltz} with $K=1$, this is equivalent to maximising the entropy:
\begin{equation}
\mathfrak{H}^{(\mu)} = \ln \mathbb{P}^{(\mu)} = \sum\limits_{j=1}^{J}  \Bigl \{ - \ln G_j  + \sum\limits_{i=1}^{I(j)} (m_{ij} \ln g_{ij} - \ln {m_{ij}!}) \Bigr \}
\label{eq:H_micro}
\end{equation}
subject to the same constraints. We {again} emphasise that \eqref{eq:H_micro} is defined on the space of climate models, and has no connection to the thermodynamic entropy $S$.  If one adopts the 
\citet{Stirling_1730} approximation $\ln {m_{ij}!}$  $\approx$ $m_{ij} \ln m_{ij}$ $- m_{ij}$ for large $m_{ij}$ (in fact this is not strictly valid), \eqref{eq:H_micro} reduces to:
\begin{equation}
\mathfrak{H}_{St}^{(\mu)}  = \sum\limits_{j=1}^{J} \Bigl \{ - \ln G_j  - \sum\limits_{i=1}^{I(j)} (m_{ij} \ln \frac{m_{ij}}{g_{ij}} + m_{ij}) \Bigr \}
\label{eq:H_micro_St}
\end{equation}
Extremisation of \eqref{eq:H_micro_St} subject to \eqref{eq:C0j_micro}-\eqref{eq:CE_micro} yields the (microcanonical) Boltzmann distribution at equilibrium:
\begin{equation}
m_{ij}^* = g_{ij} \, e^{-(\lambda_{0j}+1) - \lambda_E \epsilon_{ij}} = \frac{1}{Z_j^{(\mu)}} \, g_{ij} \, e^{- \lambda_E \epsilon_{ij}}
\label{eq:mij_St}
\end{equation}
where $*$ denotes the asymptotic (Stirling-approximate) extremum, $\lambda_{0j}$ and $\lambda_E$ are Lagrangian multipliers respectively for the allocation \eqref{eq:C0j_micro} and total energy \eqref{eq:CE_micro} constraints, and $Z_j^{(\mu)} = e^{\lambda_{0j}+1} = \sum\nolimits_{i=1}^{I(j)} g_{ij} e^{- \lambda_E \epsilon_{ij}}$ is the $j$th microcanonical partition function. Eq.\ \eqref{eq:mij_St} can be solved in conjunction with \eqref{eq:CE_micro} to calculate the predicted occupancies $m_{ij}^*$. If the occupancies are restricted to discrete values $\{0,1\}$, this will yield the choices $i(j)$ of the optimal climate model, subject to the total energy constraint $E$.  In practice, numerical solution will typically give floating-point values of $m_{ij}^*$, which can be used as weighting factors with which to combine multiple models of the same total energy $E$.  

As noted, since $m_{ij} \in \{0,1\}$, Stirling's approximation does not strictly apply to the above analysis, and so \eqref{eq:mij_St} is only an approximate solution.  This can be addressed by directly maximising the non-asymptotic entropy \eqref{eq:H_micro} with respect to $m_{ij}$, subject to \eqref{eq:C0j_micro}-\eqref{eq:CE_micro}, giving the equilibrium distribution \cite{Niven_2005, Niven_2006, Niven_2009a, Niven_2009b}:
\begin{equation}
m_{ij}^{\#} = \Lambda^{-1} \bigl( \ln g_{ij} - \lambda_{0j} - \lambda_E \epsilon_{ij} \bigr)
\label{eq:mij}
\end{equation}
where $\#$ denotes the non-asymptotic extremum and $\Lambda^{-1}(y)= \psi^{-1}(y-1)$ is the upper inverse of the function $\Lambda(x)=\psi(x+1)$, defined for convenience, in which $\psi(x)$ is the digamma function. (Note \eqref{eq:mij} can be written with additional terms in $G_j$ and $N$ \citep[c.f.][]{Niven_2009b}; these are here incorporated in $\lambda_{0j}$.) In this case, no explicit partition functions exist, and \eqref{eq:mij} must be solved in conjunction with \eqref{eq:C0j_micro}-\eqref{eq:CE_micro}.  This method will give more precise values of the optimal weighting factors $m_{ij}^{\#}$, although in practice, its numerical solution can be difficult. The non-asymptotic solution \eqref{eq:mij} is itself an approximation to the true discrete MaxProb solution (with $m_{ij} \in \{0,1\}$), which must be identified by a (computationally expensive) combinatorial search scheme.

{\it Example}: The above framework can be demonstrated by a simple example, in which a climate model is constructed from $J=3$ components, with $I=[3,4,3]$ choices of algorithm.  The degeneracies and energy levels are taken as:
\begin{gather}
\bb{g} = \left[ \begin {array}{ccc} 1&2&1\\ \noalign{\medskip}2&4&2
\\ \noalign{\medskip}4&8&4\\ \noalign{\medskip}&16&\end {array}
 \right], \qquad
\bb{\epsilon}= \left[ \begin {array}{ccc} 1&1&1\\ \noalign{\medskip}4&4&4
\\ \noalign{\medskip}9&9&9\\ \noalign{\medskip}&16&\end {array}
 \right] \text{units}
\label{eq:gv_epv}
\end{gather}
In this framework, more (degenerate) algorithms, and algorithms with a fourth energy level, are available for model component $2$. For a total energy per model of $E=17$ units, the inferred asymptotic \eqref{eq:mij_St} and non-asymptotic \eqref{eq:mij} solutions are, respectively:
\begin{gather}
\begin{split}
\bb{m}^* &= \left[ \begin {array}{ccc}  0.2823& 0.2250& 0.2823
\\ \noalign{\medskip} 0.3650& 0.2909& 0.3650
\\ \noalign{\medskip} 0.3527& 0.2811& 0.3527
\\ \noalign{\medskip}& 0.2031&\end {array} \right], 
\\
\bb{\lambda_0}^* &= \left[0.1192, 1.0393, 0.1191 \right]^{\top}, \hspace{5pt}
\lambda_E^*=0.1455
\end{split}
\label{eq:mij_St_v}
\end{gather}
and
\begin{gather}
\begin{split}
\bb{m}^{\#} &= \left[ \begin {array}{ccc}  0.1950& 0.1695& 0.1950
\\ \noalign{\medskip} 0.4206& 0.3883& 0.4206
\\ \noalign{\medskip} 0.3844& 0.3532& 0.3844
\\ \noalign{\medskip}& 0.0890&\end {array} \right], 
\\
\bb{\lambda_0}^{\#} &= \left[0.1494, 0.8755, 0.1494 \right]^{\top}, 
\hspace{5pt}
\lambda_E^{\#}=0.1460
\end{split}
\label{eq:mij_v}
\end{gather}
All calculations were conducted in Maple 14. The equilibrium model should thus be constructed using the weights in $\bb{m}^*$ (or, arguably, $\bb{m}^{\#}$). In this example, algorithms of intermediate cost (the second energy level) have the highest weighting.  Some difference is evident between the asymptotic and non-asymptotic solutions $\bb{m}$ and Massieu functions $\bb{\lambda_{0}}$, {due to the small model space of this simplified example.} The energy multipliers $\lambda_E$ of the two solutions are, however, quite similar. 

\subsection{\label{modelB} {``Canonical'' Framework I}}

The foregoing methodology is mathematically sound and provides a formal framework for the combination of different climate models.  It is, however, somewhat restrictive in that it only includes models of a single total energy $E$.  It is possible to conduct the analysis at a higher ``canonical'' level -- in the same manner as in thermodynamics -- by the analysis of ``systems of systems'', in this case involving ensembles of individual climate models.  This is shown in Figure \ref{fig:comb2}(b), in which an ensemble is constructed by collecting a sample (without replacement) of $N$ individual models, and amalgamating the results.  This can be represented by a combinatorial scheme in which distinguishable balls -- labelled by the model index $z \in \{1,...,N\}$ -- are allocated to distinguishable levels $i(j)$, again indicating choices of energy level $\epsilon_{ij}$ with degeneracy $g_{ij}$.  This gives the occupancies $n_{ij} \in \{0,\mathbb{N} \}$ for each energy level of the ensemble, which are connected to those for each model by:
\begin{gather}
n_{ij} = \sum\limits_{z=1}^N m_{ij}^{(z)}, \forall j
\\
N = \sum\limits_{i=1}^{I(j)} n_{ij} =  \sum\limits_{i=1}^{I(j)} \sum\limits_{z=1}^N m_{ij}^{(z)}, \forall j
\label{eq:C0j_can}
\end{gather}
The probability of a specified set of occupancies $n_{ij}$ for a particular $j$ is now given by \citep{Brillouin_1930, Fortet_1977, Read_1983}:
\begin{equation}
\mathbb{P}_{j}^{ (\chi)}  = \text{Prob}(\vect{n}_j|\vect{g}_j,N,\vect{\theta})
=\frac{N!}{G_j^N} \prod\limits_{i=1}^{I(j)} \frac{g_{ij}^{n_{ij}}}{n_{ij}!} 
\label{eq:Pj_can}
\end{equation}
where $\vect{n}_j = \{n_{1j},...,n_{I(j)j}\}$, while $\chi$ denotes the canonical framework. The multinomial factor $N! / \prod\nolimits_{i=1}^{I(j)} {n_{ij}!}$ accounts for number of permutations of models which attain the same set of occupancies $\vect{n}_j$. The probability of a specified ensemble, again assuming $J$ independent components, is thus given by the ``multi-multinomial'':
\begin{equation}
\begin{split}
\mathbb{P}^{ (\chi)} &= \text{Prob}(\vect{n}|\vect{g},N,\vect{\theta})  = \prod\limits_{j=1}^{J} \text{Prob}(\vect{n}_j|\vect{g}_j,N,\vect{\theta})
  \\
&= \prod\limits_{j=1}^{J} \mathbb{P}_{j}^{ (\chi)} = \prod\limits_{j=1}^{J} \frac{N!}{G_j^N} \prod\limits_{i=1}^{I(j)} \frac{g_{ij}^{n_{ij}}}{n_{ij}!} 
\label{eq:P_can}
\end{split}
\end{equation}
where $\vect{n}$ is the matrix of $n_{ij}$, whence \eqref{eq:Boltz} with $K=N^{-1}$ gives the entropy:
\begin{equation}
\begin{split}
\mathfrak{H}^{(\chi)} &= \frac{1}{N} \ln \mathbb{P}^{(\chi)}
= \frac{1}{N} \sum\limits_{j=1}^{J}  \Bigl \{ \ln N! - N \ln G_j  \\ &+ \sum\limits_{i=1}^{I(j)} (n_{ij} \ln g_{ij} - \ln {n_{ij}!}) \Bigr \}
\end{split}
\label{eq:H_can}
\end{equation}
This is subject to constraints on the occupancies, given by the first part of \eqref{eq:C0j_can}. 

How should we analyse ensembles of models? We could, in the first instance, examine the set of all possible models, of cardinality $\prod\nolimits_{j=1}^J G_j^N$ \citep{Niven_G_2009}. This would not, however, be very informative, since all models would {\it a priori} be of equal weight and so would not be discriminated by the MaxProb (or MaxEnt) method. The total ensemble also does not allow the inclusion of additional information about the desired set of models.  If, on the other hand, we impose a constraint on the mean energy of the ensemble:
\begin{gather}
\frac{1}{N} \sum\limits_{j=1}^J \sum\limits_{i=1}^{I(j)} \epsilon_{ij} n_{ij} = \langle E \rangle
\label{eq:CE_can}
\end{gather}
we then impose a decision rule on its desired composition, namely, on the average cost of constructing its constituent models.  In contrast to the microcanonical framework, this allows models of greater-than-average total energy $E> \langle E \rangle$, so long as these are balanced in the ensemble by models of lower energy $E< \langle E \rangle$. Combining \eqref{eq:H_can} with \eqref{eq:C0j_can} and \eqref{eq:CE_can} gives the Lagrangian:
\begin{equation}
\begin{split}
L^{(\chi)} &= \sum\limits_{j=1}^{J}  \Bigl \{ \frac{1}{N} \ln N! -  \ln G_j  + \sum\limits_{i=1}^{I(j)} (\frac{n_{ij}}{N} \ln g_{ij} - \frac{1}{N} \ln {n_{ij}!}) \Bigr \} \\
& - \sum\limits_{j=1}^{J} \kappa_{0j} \Bigl \{ \sum\limits_{i=1}^{I(j)} \frac{n_{ij}}{N} -1 \Bigr \} 
- \kappa_E \Bigl \{ \sum\limits_{j=1}^J \sum\limits_{i=1}^{I(j)} \epsilon_{ij} \frac{n_{ij}}{N} - \langle E \rangle \Bigr \}
\end{split}
\label{eq:L_can}
\end{equation}
where $\kappa_{0j}$ and $\kappa_E$ are Lagrangian multipliers for the allocation \eqref{eq:C0j_can} and energy \eqref{eq:CE_can} constraints. Extremisation gives the non-asymptotic equilibrium solution:
\begin{gather}
n_{ij}^{\#} = \Lambda^{-1} \bigl( \ln g_{ij} - \kappa_{0j} - \kappa_E \epsilon_{ij} \bigr), \forall j
\label{eq:nij}
\end{gather}
(again all constant terms are brought into $\kappa_{0j}$). For any given $N$, these can be solved numerically in conjunction with the constraints \eqref{eq:C0j_can} and \eqref{eq:CE_can}, to give the optimum number of times (weighting factor) $n_{ij}^{\#}$ that each choice $i(j)$ should be included in the ensemble, 
subject to $\langle E \rangle$.

When the factorials in \eqref{eq:H_can} satisfy Stirling's approximation, \eqref{eq:nij} gives the (canonical) Boltzmann distribution at equilibrium:
\begin{equation}
\frac{n_{ij}^*}{N} = \frac{1}{Z_j^{(\chi)}} \, g_{ij} \, e^{- \kappa_E \epsilon_{ij}}
\label{eq:nij_St}
\end{equation}
where $Z_j^{(\chi)} = N e^{\kappa_{0j}+1} = \sum\nolimits_{i=1}^{I(j)} g_{ij} e^{- \kappa_E \epsilon_{ij}}$ is the $j$th canonical partition function. 

{\it Example}: The canonical framework can be demonstrated using the example described previously \eqref{eq:gv_epv}, now constrained by a mean total energy per model of $\langle E \rangle =17$ units (less than the mean of all possible models, $\langle E \rangle = 24.3904$ units). The inferred asymptotic solution \eqref{eq:nij_St} is identical to \eqref{eq:mij_St_v}, i.e.:
\begin{gather}
\begin{split}
\frac{\bb{n}^*}{N} &= \bb{m}^*, \\
\bb{\kappa_0}^* &= 
\left[ \begin {array}{c} \ln  \left(  3.062388620\,{N}^{-1} \right) -
1\\ \noalign{\medskip}\ln  \left(  7.685321125\,{N}^{-1} \right) -1
\\ \noalign{\medskip}\ln  \left(  3.062388620\,{N}^{-1} \right) -1
\end {array} \right],
\hspace{5pt}
\kappa_E^*=\lambda_E^*
\end{split}
\label{eq:nij_St_v}
\end{gather}
For $N=27$ (say) this gives $\bb{\kappa_0}^* =$ $[- 3.1766$, $-2.2565$, $- 3.1766]^{\top}$. In comparison, the non-asymptotic solution \eqref{eq:nij} at $N=27$ is:
\begin{gather}
\begin{split}
\frac{\bb{n}^{\#}}{N} &= 
\left[ \begin {array}{ccc}  0.2795&
 0.2232& 0.2795\\ \noalign{\medskip}
 0.3668& 0.2940& 0.3668
\\ \noalign{\medskip} 0.3537& 0.2834&
 0.3537\\ \noalign{\medskip} & 0.1994&
 \end {array} \right], 
\\
\bb{\kappa_0}^{\#} &= \left[-2.2315,  -1.3292,  -2.2315 \right]^{\top}, 
\hspace{5pt}
\kappa_E^{\#}=0.1455
\end{split}
\label{eq:nij_v}
\end{gather}
Compared to \eqref{eq:mij_v}, the latter exhibits a more uniform distribution for each $j$, and is closer to the asymptotic form \eqref{eq:nij_St_v}.  

\subsection{\label{modelC} {``Canonical'' Framework II}}

{One difficulty with the above canonical framework is that it -- like its microcanonical precursor -- still requires separability of the model into $J$ distinct components, for which the costs $\epsilon_{ij}$ are additive.  In more general situations, this separability may not be possible due to coupling between components. In that case we must revert to a model space based on ensembles of entire models.  Severing all connection to the components $j$, we consider a model space from which we collect a sample (ensemble) of $\II{N}$ models, containing ${n}_\imath$ models each of total energy $E_\imath$. Each energy level has degeneracy $g_i$.  The probability of a specified ensemble is: }
{
\begin{equation}
\begin{split}
\mathbb{P}^{(\chi)}_{II}  = \text{Prob}(\vect{n}|\vect{g},\II{N},\vect{\theta})
&= \frac{\II{N}!}{\II{G}^\II{N}} \prod\limits_{\imath=1}^{I} \frac{g_{\imath}^{n_{\imath}}}{n_{\imath}!} 
\label{eq:P_canII}
\end{split}
\end{equation}
where $\II{G} = \sum\nolimits_{\imath=1}^{I} g_{\imath}$. Boltzmann's equation \eqref{eq:Boltz} with $K=\II{N}^{-1}$ gives the entropy:
\begin{equation}
\begin{split}
\mathfrak{H}^{(\chi)}_{II} &= \frac{1}{\II{N}} \ln \mathbb{P}^{(\chi)}_{II}
= \frac{1}{\II{N}} \Bigl \{ \ln \II{N}! - \II{N} \ln \II{G}  \\ &+  \sum\limits_{\imath=1}^{I}   (n_{\imath} \ln g_{\imath} - \ln {n_{\imath}!}) \Bigr \}
\end{split}
\label{eq:H_canII}
\end{equation}
This is subject to the occupancy and mean ensemble energy constraints:
\begin{gather}
\sum\limits_{\imath=1}^{I} n_{\imath} = \II{N} 
\label{eq:CN_canII} 
\\
\frac{1}{\II{N}} \sum\limits_{\imath=1}^I E_{\imath} n_{\imath} = \langle E \rangle
\label{eq:CE_canII}
\end{gather}
Forming the Lagrangian and extremisation gives the non-asymptotic equilibrium occupancies:
\begin{gather}
n_{\imath}^{\#} = \Lambda^{-1} \bigl( \ln g_{\imath} - \varphi_{0} - \varphi_E E_{\imath} \bigr) 
\label{eq:ni_canII}
\end{gather}
where $\varphi_{0}$ and $\varphi_E$ are Lagrangian multipliers for the occupancy and energy constraints. If \eqref{eq:H_canII} satisfies the Stirling approximation, the distribution reduces to:
\begin{equation}
\frac{n_{\imath}^*}{\II{N}} = \frac{1}{Z^{(\chi)}_{II} } \, g_{\imath} \, e^{- \varphi_E E_{\imath}}
\label{eq:ni_St_canII}
\end{equation}
where $Z^{(\chi)}_{II}= \II{N} e^{\varphi_{0}+1} = \sum\nolimits_{\imath=1}^{I} g_{\imath} e^{- \varphi_E E_{\imath}}$ is the canonical II partition function. Either \eqref{eq:ni_canII} or (if valid) \eqref{eq:ni_St_canII} can be solved in conjunction with the constraints \eqref{eq:CN_canII}-\eqref{eq:CE_canII}, to give the weights $n_\imath$ of the most representative model. }

\subsection{\label{summary} Summary}

At this point, it is worth summarising some important features of the microcanonical and {two} canonical frameworks proposed:
\begin{list}{$\bullet$}{\topsep 3pt \itemsep 3pt \parsep 0pt \leftmargin 8pt \rightmargin 0pt \listparindent 0pt
\itemindent 0pt}
\item As evident from the predicted solutions \eqref{eq:mij_St}-\eqref{eq:mij} and \eqref{eq:nij}-\eqref{eq:nij_St}, if one seeks the optimal model to describe a set of climate models, it is not necessary to compute all possible combinations of models. Using the MaxProb method, one can {\it directly} calculate the single model or a reduced set of models which best represents the model set, subject to constraint(s) on the model or ensemble properties. The effect of two competing constraints is examined in the next section.
\item The microcanonical framework imposes constraint(s) on individual models, whereas the {two} canonical frameworks impose constraint(s) over ensembles of models.  The latter enable the synthesis of larger sets of models. 
\item Note that, due to the assumed independence of the $J$ model components, the {microcanonical and canonical I frameworks} are ``multi-multinomial'' \eqref{eq:P_micro} and \eqref{eq:P_can}. The choices $i(j)$ for a specified $j=\vartheta$ are thus independent of the other choices $j \ne \vartheta$. {The MaxProb prediction can therefore} be computed using individual models composed of whichever choices $i(j)$ are convenient, so long as the overall set conforms to the MaxProb prediction.  {In the canonical II model, we overcome the difficulty of coupled model components by considering ensembles of entire models, with constraints on the total energy of each model.}
\item How should we interpret the Lagrangian multipliers on the energy constraint?  By analogy with thermodynamics, these can be interpreted as $\lambda_E = 1/kT^{(\mu)}$, $\kappa_E = 1/kT^{(\chi)}$ {and $\varphi_E = 1/kT^{(\chi)}_{II}$, where the $T$ parameters are framework ``temperatures''} and $k$ is a constant with units of energy (or cost) per temperature unit. {The $T$'s are not thermodynamic temperatures, but express the distribution of energy over the available energy levels, in the relevant model or ensemble space. In effect, they} serve as proxy variables for the total model cost $E$ or mean ensemble cost $\langle E \rangle$.    
\item Although the MaxProb framework is primarily designed to determine the most probable (maximum entropy) model, it is also possible to interrogate the Lagrangian to determine the minimum entropy model(s), i.e.\ those which lie farthest from the optimum.  In this manner, one can explore the extremities of the model or ensemble space, to identify model outliers. Since minimum entropy solutions tend to lie on non-continuous boundaries of the solution domain, they are generally inaccessible to extremisation methods \citep{Kapur_K_1992}; nonetheless, they should be identifiable by numerical optimisation algorithms such as simulated annealing. 
\item The mathematical structure of the output from the MaxEnt algorithm gives rise to many more features of the predicted solution.  Some of these features are explored in later sections.
\end{list}

\section{\label{modelC} ``Canonical'' Framework {II} with Cost and Benefit Constraints}
\subsection{\label{modelCderiv}Derivation}

We now examine a more comprehensive canonical {II} framework, in which we impose two constraints at the ensemble level: a constraint on the mean ensemble cost or energy $\langle E \rangle$, as before \eqref{eq:CE_canII}, and also a constraint on some measure of the average ensemble ``worthiness'' or ``benefit'' $\langle B \rangle$ (for example, a measure of its precision or accuracy). In this manner, we construct a MaxProb framework with which to conduct cost-benefit analyses of various ensembles of models, and to interrogate the trade-off between costs and benefits\footnote{This approach is applicable not only to climate models, but models of any type, including economic models.}.  In general, the energy and benefit levels will have different ranks, necessitating the use of different indices {$i \in \{1, ..., I\}$ (as before) and $\ell \in \{1, ..., \ELL\}$. We therefore consider model choices ranked by total model energies $E_{i\ell}$ and benefits $B_{i\ell}$, of joint degeneracy $g_{i \ell}$.  The probability that an ensemble of $\II{N}$ models has the occupancies $\{ n_{i \ell} \}$ is} governed by the multinomial:
\begin{equation}
\mathbb{P} = \frac{\II{N}!}{\II{G}^\II{N}} \prod\limits_{i=1}^{I} \prod\limits_{\ell=1}^{\ELL}  \frac{g_{i\ell}^{n_{i\ell}}}{n_{i\ell}!} 
\label{eq:P_can2}
\end{equation}
where now $\II{G} = \sum\nolimits_{i=1}^{I} \sum\nolimits_{\ell=1}^{\ELL} g_{i\ell}$ (for convenience we drop the super- and subscript labels).  From \eqref{eq:Boltz} with $K=\II{N}^{-1}$, we maximise the entropy: 
\begin{equation}
\begin{split}
\mathfrak{H} &= \frac{1}{\II{N}}  \Bigl \{ \ln \II{N}! - \II{N} \ln \II{G}  \\
& \hspace{30pt} + \sum\limits_{i=1}^{I} \sum\limits_{\ell=1}^{\ELL} (n_{i\ell} \ln g_{i\ell} - \ln {n_{i\ell}!}) \Bigr \}
\end{split}
\label{eq:H_can2}
\end{equation}
subject to the constraints:
\begin{gather}
\frac{1}{\II{N}} \sum\limits_{i=1}^{I} \sum\limits_{\ell=1}^{\ELL} n_{i\ell} = 1
\label{eq:C0_can2}
\\
\frac{1}{\II{N}} \sum\limits_{i=1}^{I} \sum\limits_{\ell=1}^{\ELL} E_{i\ell} n_{i\ell} = \langle E \rangle
\label{eq:CE_can2} 
\\
\frac{1}{\II{N}} \sum\limits_{i=1}^{I} \sum\limits_{\ell=1}^{\ELL} B_{i\ell} n_{i\ell} = \langle B \rangle
\label{eq:CB_can2}
\end{gather}
to give the non-asymptotic equilibrium solution:
\begin{gather}
n_{i\ell}^{\#} = \Lambda^{-1} \bigl( \ln g_{i\ell} - \omega_{0} - \omega_E E_{i\ell} - \omega_B B_{i\ell} \bigr)
\label{eq:nil2}
\end{gather}
where $\omega_{0}$, $\omega_E$ and $\omega_B$ are the Lagrangian multipliers. If Stirling's approximation applies, the entropy is:
\begin{equation}
\begin{split}
\mathfrak{H}_{St} =  \Bigl \{ \ln \II{N}  -  \ln \II{G}  
- \sum\limits_{i=1}^{I} \sum\limits_{\ell=1}^{\ELL} (\frac{n_{i\ell}}{\II{N}}  \ln \frac{n_{i\ell}}{g_{i\ell}} ) \Bigr \}
\end{split}
\label{eq:H_can2_St}
\end{equation}
whence extremisation gives:
\begin{equation}
\begin{split}
\frac{n_{i\ell}^*}{\II{N}} 
&=  \frac{ g_{i\ell}}{Z}  \, e^{- \omega_E E_{i\ell}  - \omega_B B_{i\ell}} 
\\
Z &= \II{N} e^{\omega_{0}+1} = \sum\limits_{i=1}^{I} \sum\limits_{\ell=1}^{\ELL} g_{i \ell} \, e^{- \omega_E E_{i\ell}  - \omega_B B_{i\ell}}
\label{eq:nil2_St}
\end{split}
\end{equation}
where $Z$ is the partition function. 

The Lagrangian multiplier $\omega_E$ can again be interpreted as an inverse ensemble temperature $\omega_E = 1/kT$, where $k$ is a constant. The multiplier $\omega_B$ can be considered as a measure of the overall benefit provided by the ensemble, in reciprocal benefit units. In effect, it acts as a proxy variable for the mean benefit $\langle B \rangle$.  Since $\langle B \rangle$ measures the average information or value provided by the framework, it can be interpreted very crudely as a reciprocal density or volume, whereupon we can interpret $\omega_B = P/kT$, in which $P$ is a mean ensemble pressure (this interpretation should not be taken too seriously).

\subsection{\label{Jaynes}Jaynesian Mathematical Structure}

Now that we have the main results, we can examine several important mathematical features of the solution. Most of these were reported in a generic context by 
\citet{Jaynes_1957, Jaynes_1963, Jaynes_2003} (see also \citet{Kapur_K_1992} and \citet{Tribus_1961b}), although many were previously known in thermodynamics.  The foregoing microcanonical and canonical {I and II} frameworks also exhibit these features, but it is more interesting to examine the effect of two competing constraints.

Firstly, {for the Stirling-approximate case, substitution of \eqref{eq:nil2_St} into \eqref{eq:H_can2_St}, by sorting into expectations along the lines of \citep{Jaynes_1963}, gives the asymptotic maximum entropy:}
{
\begin{align}
\begin{split}
\mathfrak{H}^{*} &=  -  \ln \II{G} - \phi  + \omega_E \langle E \rangle + \omega_B \langle B \rangle 
\end{split}
\label{eq:H_can2_St_max}
\end{align}}
{where for convenience we define the potential function (negative Massieu function) $\phi = -\omega_{0} = -\ln Z$.  The most probable state of the ensemble is thus given by a constant term, plus the Massieu function}, plus the sum of products of the constraints and their conjugate Lagrangian multipliers.   

Since the entropy function and constraints are state variables on the space of ensembles of models, \eqref{eq:H_can2_St_max} provides a linear homogenous equation which describes the framework\footnote{{Strictly, if the initial terms in $G$ and $\phi$ are constant, the differential of \eqref{eq:H_can2_St_max} is a linear homogeneous first-order differential equation. Absorbing the constant into $\phi$, \eqref{eq:H_can2_St_max} can then be interpreted as an Euler equation \citep[c.f.][]{Callen_1985}.}}.   This can be used to examine the response of the framework to changes in the constraints and/or multipliers. For constant {$G$ and $\phi$} we 
immediately see that \citep{Jaynes_1957, Jaynes_1963, Jaynes_2003}:
\begin{equation}
\frac {\partial \mathfrak{H}^{*}}{\partial \langle E \rangle} \biggr|_{\langle B \rangle} = \omega_E, \hspace{30pt}  \frac {\partial \mathfrak{H}^{*}}{\partial \langle B \rangle} \biggr|_{\langle E \rangle} = \omega_B  
\label{eq:H_diff_rels}
\end{equation}
Second differentiation gives the Hessian matrix: 
\begin{equation}
- \tens{a} =
\begin{bmatrix} 
\dfrac{\partial^2 \mathfrak{H}^{*}}{\partial \langle E \rangle^2} 
&\dfrac {\partial^2 \mathfrak{H}^{*}}{\partial \langle E \rangle \partial \langle B \rangle} 
\\\\
\dfrac {\partial^2 \mathfrak{H}^{*}}{\partial \langle B \rangle \partial \langle E \rangle} 
&\dfrac{\partial^2 \mathfrak{H}^{*}}{\partial \langle B \rangle^2}
\end{bmatrix} 
= 
\begin{bmatrix} 
\dfrac{\partial \omega_E}{\partial \langle E \rangle}
&\dfrac{\partial \omega_B  }{\partial \langle E \rangle} 
\\\\
\dfrac{\partial \omega_E}{\partial \langle B \rangle} 
&\dfrac{\partial \omega_B  }{\partial \langle B \rangle}
\end{bmatrix} 
\label{eq:H_diff2_rels}
\end{equation}
If the mixed derivatives are equivalent (i.e. $\mathfrak{H}^{*}$ is continuous and continuously differentiable, at least up to second order), this gives the reciprocal or Maxwell-like relation \citep{Jaynes_1963, Jaynes_2003}:
\begin{equation}
\frac{\partial \omega_E}{\partial \langle B \rangle} = \frac{\partial \omega_B  }{\partial \langle E \rangle}
\label{eq:H_Maxwell}
\end{equation}
Equivalently, \eqref{eq:H_can2_St_max} can be rewritten as a function of the potential $\phi$, whence it can be shown that \citep{Jaynes_1957, Jaynes_1963, Jaynes_2003}:
\begin{equation}
\frac {\partial \phi}{\partial \omega_E} \biggr|_{\omega_B} = \langle E \rangle, \hspace{30pt}  
\frac {\partial \phi}{\partial \omega_B} \biggr|_{\omega_E} = \langle B \rangle  
\label{eq:phi_diff_rels}
\end{equation}
Second differentiation gives: 
\begin{equation}
- \vect{\alpha} =
\begin{bmatrix} 
\dfrac{\partial^2 \phi}{\partial \omega_E^2} 
&\dfrac {\partial^2 \phi}{\partial \omega_E \partial \omega_B} 
\\\\
\dfrac {\partial^2 \phi}{\partial \omega_B \partial \omega_E} 
&\dfrac{\partial^2 \phi}{\partial \omega_B^2}
\end{bmatrix} 
= 
\begin{bmatrix} 
\dfrac{\partial \langle E \rangle}{\partial \omega_E}
&\dfrac{\partial \langle B \rangle}{\partial \omega_E} 
\\\\
\dfrac{\partial \langle E \rangle}{\partial \omega_B  } 
&\dfrac{\partial \langle B \rangle}{\partial \omega_B  }
\end{bmatrix} 
\label{eq:phi_diff2_rels}
\end{equation}
giving, again for equivalent mixed derivatives \citep{Jaynes_1963, Jaynes_2003}:
\begin{equation}
\frac{\partial \langle B \rangle}{\partial \omega_E} = \frac{\partial \langle E \rangle}{\partial \omega_B  }
\label{eq:phi_Maxwell}
\end{equation}
From \eqref{eq:H_diff2_rels} and \eqref{eq:phi_diff2_rels}, it is evident that:
\begin{equation}
\tens{a} = \vect{\alpha}^{-1}
\label{eq:Legendre}
\end{equation}
This defines a Legendre transformation between $\mathfrak{H}^{*}$ and $\phi$ representations of the system \citep{Jaynes_1963, Jaynes_2003, Kapur_K_1992}.    

{Finally, we note that it may be desirable to rank climate models by more than two  properties, e.g. the model cost $E$ and several different benefits $B_1, B_2, ..., B_M$.  The foregoing analysis can readily be extended into as many dimensions as desired, giving the above mathematical structure as a function of the constraints $\langle E \rangle$, and $\langle B_1 \rangle$, ..., $\langle B_M \rangle$. } 

\subsection{\label{modelCimp}Implications}

\begin{figure}[t] 
\setlength{\unitlength}{0.6pt}
  \begin{picture}(410,200)
   \put(5,0){\includegraphics[width=80mm]{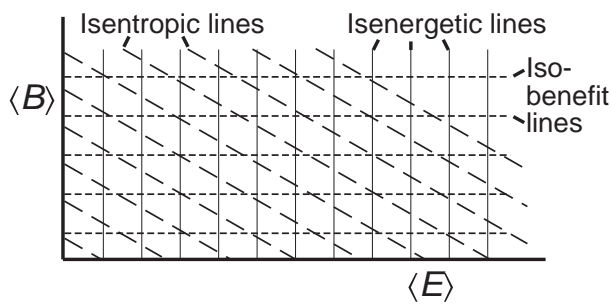} }
  \end{picture}
\caption{Schematic diagram of Gibbs' geometry, for the MaxEnt cost-benefit climate model weighting framework of \S\ref{modelC}.}
\label{fig:Gibbs}
\end{figure}

What are the implications of the above Jaynesian mathematical structure?  In essence, it governs the effect of changes to the constraints and/or multipliers on the manifold of equilibrium positions of the framework. This includes:  
\begin{list}{$\bullet$}{\topsep 3pt \itemsep 3pt \parsep 0pt \leftmargin 8pt \rightmargin 0pt \listparindent 0pt
\itemindent 0pt}
\item Firstly, the first derivatives \eqref{eq:H_diff_rels} and \eqref{eq:phi_diff_rels} can be interpreted as {\it equations of state} on the space of ensembles of models, describing the relationship between the rate of change of the entropy or potential as a function of the constraints or their conjugate multipliers \citep{Jackson_1968}.  
\item Secondly, the second derivatives \eqref{eq:H_diff2_rels} and \eqref{eq:phi_diff2_rels} describe the {\it susceptibilities} of the framework, i.e.\ the functional connections between the constraints and multipliers.  In thermodynamics, such susceptibilities include the heat capacity, isothermal compressibility, coefficient of thermal expansion and so on \citep[e.g.][]{Niven_Andresen_2009, Niven_2009c, Callen_1985, Gilmore_1982}; if desired, such parameters can also be defined for the model framework proposed here.  The Maxwell-like relations \eqref{eq:H_Maxwell} and \eqref{eq:phi_Maxwell} reflect the coupling between the constraints, such that changes in one constraint or its multiplier, at constant $\mathfrak{H}^{*}$ or $\phi$, will produce adjustments to the other pair.  
\item Thirdly, the second derivative matrix \eqref{eq:phi_diff2_rels} of the potential function $\phi$ contains even further information, since in the asymptotic limit ($N \to \infty$), it is equivalent (with change of sign) to the variance-covariance matrix of the constraints \citep{Jaynes_1957, Jaynes_1963, Jaynes_2003, Kapur_K_1992}:
\begin{equation}
\vect{\alpha} =
\begin{bmatrix} 
\langle E^2 \rangle - \langle E \rangle^2, 
&\langle E B \rangle - \langle E \rangle \langle B \rangle
\\\\
\langle E B \rangle - \langle E \rangle \langle B \rangle,
&\langle B^2 \rangle - \langle B \rangle^2 
\end{bmatrix} 
\label{eq:phi_cov}
\end{equation}
{Accordingly, $\vect{\alpha}$ is positive definite (or semi-definite if singularities exist) \citep{Kapur_K_1992}. From the Legendre transformation \eqref{eq:Legendre}, $\tens{a}$ is also positive definite (or semi-definite) \citep{Kapur_K_1992}.}  In consequence, from \eqref{eq:H_diff2_rels} and \eqref{eq:phi_diff2_rels} (including the tensor sign reversals), $\mathfrak{H}^{*}(\langle E \rangle, \langle B \rangle)$ and $\phi(\omega_E, \omega_B)$ are both concave functions.  Furthermore, the diagonal of \eqref{eq:phi_cov} gives the magnitude of the standard deviation or {\it ``fluctuations''} of the ensemble with respect to each constraint, usually expressed in normalised form by the coefficients of variation \citep{Callen_1985}:
\begin{equation}
\begin{split}
C_V(E) = \frac{\sqrt{\langle E^2 \rangle - \langle E \rangle^2}}{\langle E \rangle} = \frac{1}{\langle E \rangle}  \sqrt{-\dfrac{\partial \langle E \rangle}{\partial \omega_E}}
\\
C_V(B) = \frac{\sqrt{\langle B^2 \rangle - \langle B \rangle^2}}{\langle B \rangle} = \frac{1}{\langle B \rangle} \sqrt{-\dfrac{\partial \langle B \rangle}{\partial \omega_B  }}
\end{split}
\label{eq:phi_CV}
\end{equation}
The covariance, similarly normalised, provides a measure of the coupling between constraints \citep{Callen_1985}:
\begin{equation}
\begin{split}
&C_V(E,B) = \sqrt{\frac{ \langle E B \rangle - \langle E \rangle \langle B \rangle }{\langle E \rangle \langle B \rangle}} 
\\
& \hspace{5pt} = \sqrt{\frac{1}{\langle E \rangle \langle B \rangle} \biggl | - \frac{\partial \langle B \rangle}{\partial \omega_E} \biggr | }  
= \sqrt{\frac{1}{\langle E \rangle \langle B \rangle} \biggl | - \frac{\partial \langle E \rangle}{\partial \omega_B} \biggr | } 
\end{split}
\label{eq:phi_CVmixed}
\end{equation}
\item Fourthly, the manifold of predicted equilibrium positions defined by $\mathfrak{H}^{*}(\langle E \rangle, \langle B \rangle)$ or $\phi(\omega_E, \omega_B)$ can be interpreted as a {\it framework geometry}, analogous to the thermodynamic geometry examined by \citet{Gibbs_1873a, Gibbs_1873b, Gibbs_1875} (see also \citep{Callen_1985, Gaggioli_etal_2002a, Gaggioli_etal_2002b}).  For example, if we consider $\langle B \rangle$ as a function of $\langle E \rangle$, as shown graphically in Figure \ref{fig:Gibbs}, we can represent positions of constant entropy $\mathfrak{H}^*$ by a series of {\it isentropic curves} on this graph.  From \eqref{eq:H_can2_St_max}, these will be straight lines with negative gradient $-\omega_E/\omega_B$, indicating that an increase in the energy or cost $\langle E \rangle$, at constant $\mathfrak{H}^{*}$ (and $\phi$), causes a corresponding decrease in $\langle B \rangle$.  
%
%
Of course, many other curves can also be plotted on the diagram, including isoenergetic, isobenefit, iso-$\omega_E$ and iso-$\omega_B$ curves, defined by rearrangements of \eqref{eq:H_can2_St_max}.  We can also plot $\omega_B$ as a function of $\omega_E$, on which we can construct {\it isopotential curves} with negative gradient $-{\langle E \rangle}/{\langle B \rangle}$.  (Adopting the crude analogy of \S\ref{modelCderiv}, these can be transformed to plots of $P$ as a function of $T$ for the model framework.)
Three-dimensional graphs such as $\mathfrak{H}^{*}(\langle E \rangle, \langle B \rangle)$ or $\phi(\omega_E, \omega_B)$ can also be constructed, containing isosurfaces of various kinds \citep{Gibbs_1873b, Gibbs_1875}. As pointed out by \citet{Gibbs_1873a, Gibbs_1873b, Gibbs_1875}, it is advantageous to plot ``fundamental equations'' such as $\mathfrak{H}^{*}(\langle E \rangle, \langle B \rangle)$ or $\phi(\omega_E, \omega_B)$, rather than forms unobtainable from these by Legendre transformation (such as $\mathfrak{H}^{*}(\omega_E, \omega_B)$), so that all parameters not represented on the axes can be calculated for a given path simply by differentiation. 

\hspace{10pt} {Recalling that the frameworks herein consist of {\it all possible models} consistent with the constraints, the resulting manifold $\mathfrak{H}^{*}(\langle E \rangle, \langle B \rangle)$ or $\phi(\omega_E, \omega_B)$ should for the most part be continuous in its geometric space, reflecting infinitesimal changes in parameters and incremental changes in model algorithms. However, in some circumstances there may be discontinuities in the manifold, due to abrupt changes in model algorithm or adoption of different scientific paradigms.  Such changes can be described as {\it phase changes} or {\it tipping points} within the model space, leading to assortments of stable and unstable solutions and path-dependent hysteresis effects.  These may create particular difficulties, but can of course be handled in much the same manner as in thermodynamics. }
\item Finally, it can be shown that either framework $\mathfrak{H}^{*}(\langle E \rangle, \langle B \rangle)$ or $\phi(\omega_E, \omega_B)$ can be endowed with a {\it Riemannian geometry} (entirely distinct from the framework geometry just described), using the metric furnished by the respective (positive definite) Hessian matrix $\tens{a}$ or $\vect{\alpha}$ \cite{Weinhold_1975c, Ruppeiner_1979, Salamon_A_G_B_1980, Salamon_B_1983, 
Nulton_etal_1985, 
Niven_Andresen_2009}. As noted, the two metrics and hence the geometries are connected by Legendre transformation \eqref{eq:Legendre}. The Riemannian interpretation leads to an important physical limit: a {\it least action bound} on the cost, in units of $\mathfrak{H}^*$ or $\phi$, to move the framework from one equilibrium position to another at finite rates of change of the constraints or multipliers.  This bound -- which constitutes an extension of finite time thermodynamics \cite{Weinhold_1975c, Ruppeiner_1979, Salamon_A_G_B_1980, Salamon_B_1983, 
Nulton_etal_1985, 
Niven_Andresen_2009}, but is in some sense allied to the informational limits identified by \citet{Szilard_1929}, \citet{Landauer_1961}, \citet{Bennett_1973} and similar workers \citep{Leff_Rex_1990} -- is examined in more detail in Appendix A. 
\end{list} 

\section{Conclusions}
In this study, {several} maximum-entropy frameworks are presented for the synthesis of outputs from multiple Earth climate models, based on constraints on the properties of individual models (microcanonical framework) or ensembles of models {(two canonical frameworks)}. The asymptotic and non-asymptotic entropy functions for each case are derived by combinatorial reasoning, and applied to simple systems constrained by the total model energy $E$ {(microcanonical)} or mean ensemble energy $\langle E \rangle$ {(canonical)}. {In each case it is shown that the MaxEnt method identifies the most representative (most probable) model from a set of climate models, subject to the specified constraints, eliminating the need to calculate the entire set.}  The parametric and geometric implications of the underlying Jaynesian mathematical structure are examined, with reference to a canonical framework with competing cost and benefit constraints, allowing interrogation of the trade-off between costs and benefits.  Finally, a finite-time limit on the minimum cost of modification of the synthesis framework, at finite rates of change, is also reported. 

{The foregoing analysis therefore provides climate modellers -- or those who must rank and combine climate models -- with a rational tool to amalgamate a large set of models into a single representative model (or a small representative set). This enables the weighting of climate projections from different groups, and will also dramatically reduce the computational demand on the climate modelling community.  Indeed, the benefits extend into other fields: as commented by a reviewer, for long-range weather forecasts it is common practice to combine projections from different meteorological models, to improve reliability. The MaxEnt frameworks proposed here could equally be applied to this task.}

A caveat to the foregoing analysis is that the inferred equilibrium climate model will not necessarily be the ``most correct'' model, but merely the one which is most representative of the available set of models.  If the model space is incomplete, or their underlying physical or modelling assumptions are incorrect, any resulting errors will also be incorporated in the equilibrium model.  A more comprehensive probabilistic framework, which incorporates the errors associated with our lack of knowledge (of data, phenomena and models), would consist of a Bayesian inferential framework extending back to all raw climate data, a substantial endeavour which -- as its minimum condition -- would require climate scientists to abandon their use of orthodox methods for statistical inference and parameter estimation \cite{Jaynes_2003}.


\begin{acknowledgements}
This work was inspired by discussions at the Mathematical and Statistical Approaches to Climate Modelling and Prediction workshop, Isaac Newton Institute for Mathematical Sciences, Cambridge, UK, 11 Aug. to 22 Dec. 2010. The author sincerely thanks the workshop organisers for travel support.

\end{acknowledgements}

\section*{Appendix A: The Least Action Bound}
The Riemannian geometric interpretation in \S\ref{modelCimp} leads to a rather curious physical limit.  Consider a path on the manifold of equilibrium positions defined by $\protect{\mathfrak{H}^{*}(\langle E \rangle, \langle B \rangle)}$ or $\protect{\phi(\omega_E, \omega_B)}$, specified by some path parameter $\xi$ in the model space, which may -- but need not -- correspond to time. The arc length of the path from position 1 to 2, represented by $\xi=0$ to $\xi = \xi_{max}$, is given by \citep{Niven_Andresen_2009}:
\begin{align}
\L = \int\limits_0^{\xi_{max}} \sqrt{  \vect{\dot{f}}^{\top} \, \tens{a} \, \vect{\dot{f}}}  \, d\xi 
= \int\limits_0^{\xi_{max}} \sqrt{  \vect{\dot{\Omega}}^{\top} \, \vect{\alpha} \, \vect{\dot{\Omega}}}  \, d\xi 
\label{eq:arclength} 
\end{align}
where $\vect{f}=[\langle E \rangle, \langle B \rangle]^{\top}$, $\vect{\Omega}=[\omega_E, \omega_B]^{\top}$ and the overdot indicates the rate of change with respect to $\xi$. Now, in the $\mathfrak{H}^*$ representation, the total change in the framework entropy along the same path can be shown to be \citep{Niven_Andresen_2009}:
\begin{align}
\Delta \mathfrak{H}^* &
=  \int\limits_{\mathfrak{H}^*_1}^{\mathfrak{H}^*_2} d \mathfrak{H}^*
= \bar{\epsilon} \int\limits_0^{\xi_{max}} \frac{1}{2} \; {  \vect{\dot{f}}^{\top} \, \tens{a} \, \vect{\dot{f}}}  \, d\xi 
= \bar{\epsilon} \II{J}
\label{eq:deltaH_tot_int}
\end{align}
where $\bar{\epsilon}$ is a mean dissipation parameter (e.g. minimum dissipation time) and $\II{J}$ is an action integral defined within the model space.  Similarly, in the $\phi$ representation, the total change is:
\begin{align}
-\Delta \phi &
=  -\int\limits_{\phi_1}^{\phi_2} d \phi
= \bar{\epsilon} \int\limits_0^{\xi_{max}} \frac{1}{2} \; {  \vect{\dot{\Omega}}^{\top} \, \vect{\alpha} \, \vect{\dot{\Omega}}}  \, d\xi 
= \bar{\epsilon} \II{J}
\label{eq:deltaphi_tot_int}
\end{align}
From the Cauchy-Schwarz inequality, \eqref{eq:arclength}-\eqref{eq:deltaphi_tot_int} give, in either case:
\begin{equation}
\II{J} \ge \frac {\L^2}{2 \, \xi_{max}}  
\label{eq:leastaction}
\end{equation}
Eq.\ \eqref{eq:leastaction} can be considered as a generalised {\it least action bound} on processes on the manifold of optimal solutions. In essence, it specifies the minimum cost or penalty, in units of $\mathfrak{H}^*$ or $\phi$, to move the system from $\xi=0$ to $\xi=\xi_{max}$ at the specified rates $\vect{\dot{f}}$ or $\vect{\dot{\Omega}}$.  If the process occurs infinitely slowly, the lower bound of the action is zero (it is ``reversible''); otherwise, it is necessary to pay the minimum penalty $\Delta \mathfrak{H}^*_{min} = - \Delta \phi_{min} = \bar{\epsilon} \II{J}_{min} = \frac{1}{2} \bar{\epsilon} {\L^2}/{\xi_{max}}$ to be able to alter the framework within the finite parameter duration $\xi_{max}$ (it is ``dissipative''). In the present scenario, we assume that the costs $\langle E \rangle$ and benefits $\langle B \rangle$ of the model framework are realisable as external physical quantities, outside the model space itself; likewise, so will be the entropy $\mathfrak{H}^*$ and potential $\phi$, either in the units of $k$ or the equivalent information units. Eq.\ \eqref{eq:leastaction} therefore provides an {\it information limit} on the minimum price for making alterations to a constrained modelling framework. (Of course, it applies to any modelling framework, not just for climate modelling.) In some sense, this limit is allied to the informational principles demonstrated by \citet{Szilard_1929}, \citet{Landauer_1961}, \citet{Bennett_1973} and many others \citep{Leff_Rex_1990}, although it is of quite different character.


\end{document}